# End-to-End Deep Learning for Real-Time Neuroimaging-Based Assessment of Bimanual Motor Skills

Aseem Subedi,[1] Rahul,[1,*] Lora Cavuoto,[2] Steven Schwaitzberg,[3] Matthew Hackett,[4] Jack Norfleet,[4] and Suvranu De[1,5]

[1] Center for Modeling, Simulation and Imaging in Medicine, Rensselaer Polytechnic Institute, Troy NY 12180, USA
[2] Department of Industrial and Systems Engineering, University at Buffalo-State University of New York, Buffalo NY 14215, USA
[3] Department of Surgery, University at Buffalo-State University of New York, Buffalo NY 14215, USA
[4] U.S. Army Futures Command, Combat Capabilities Development Command Soldier Center STTC, Orlando FL 32826, USA
[5] Florida A&M University-Florida State University College of Engineering, Tallahassee FL 32310, USA

## Abstract

The real-time assessment of complex motor skills presents a challenge in fields such as surgical training and rehabilitation. Recent advancements in neuroimaging, particularly functional near-infrared spectroscopy (fNIRS), have enabled objective assessment of such skills with high accuracy. However, these techniques are hindered by extensive preprocessing requirements to extract neural biomarkers. This study presents a novel end-to-end deep learning framework that processes raw fNIRS signals directly, eliminating the need for intermediate preprocessing steps. The model was evaluated on datasets from three distinct bimanual motor tasks—suturing, pattern cutting, and endotracheal intubation (ETI)—using performance metrics derived from both training and retention datasets. It achieved a mean classification accuracy of 93.9% (SD 4.4) and a generalization accuracy of 92.6% (SD 1.9) on unseen skill retention datasets, with a leave-one-subject-out cross-validation yielding an accuracy of 94.1% (SD 3.6). Contralateral prefrontal cortex activations exhibited task-specific discriminative power, while motor cortex activations consistently contributed to accurate classification. The model also demonstrated resilience to neurovascular coupling saturation caused by extended task sessions, maintaining robust performance across trials. Comparative analysis confirms that the end-to-end model performs on par with or surpasses baseline models optimized for fully processed fNIRS data, with statistically similar ($p<0.05$) or improved prediction accuracies. By eliminating the need for extensive signal preprocessing, this work provides a foundation for real-time, non-invasive assessment of bimanual motor skills in medical training environments, with potential applications in robotics, rehabilitation, and sports.

## 1. Introduction

Learning fine motor skills involves the development of neural pathways that enable bimanual motor coordination necessary to perform complex motor tasks, such as those required in minimally invasive surgery and emergency medicine tasks. Recently, neuroimaging techniques, such as functional near infrared spectroscopy (fNIRS), have been shown as feasible methods to assess surgical skills with high accuracy [Nemani 2018 2019, Gao 2022, Singh 2018], elucidating its potential for objective assessment of bimanual motor skills. However, the adoption of these techniques is often hindered by the need for extensive signal processing steps to extract neuronal activation data devoid of extracerebral signals (Figure 1). This work presents an end-to-end deep learning framework for the real-time assessment of bimanual motor skills directly from raw neuroimaging signals, facilitating practical field implementation for skill assessment in surgical and emergency medicine. This framework is also adaptable to real-time assessment related to neurodegenerative disorders and targeted rehabilitation interventions.

* Corresponding author, Tel: +1 (518) 276-6707; Email: rahul@rpi.edu

Simulation-based medical training paradigms, particularly those focused on bi-manual motor tasks, typically involve repeated practice, with feedback provided by human raters. However, conventional human assessment is vulnerable to various factors, including preferential technique biases and halo effects [Levin]. Objective, automated assessment can instead be achieved by tracking behavioral or physiological/cognitive responses of trainees performing certification tasks that require bimanual expertise. Related studies have used behavioral markers, such as tool movements [Estrada, Hofstad, Yanik 2023], hand movements [Watson], and eye tracking [Taha Abu] to infer proficiency. Physiological responses, like heart rate, have also been used to monitor training stress, while cognitive responses during laparoscopic tasks have been explored through methods such as pupillometry [Nguyen], Electro-Encephalography (EEG) [Guru, Shahbazi, Zhou, Shafiei, Borghini], and functional near-infrared spectroscopy (fNIRS) [Nemani 2018 2019, Gao 2022, Singh]. Among these techniques, neuroimaging techniques have been shown to be both objective and accurate [Keles] and provide a direct view of the neural changes associated with skill learning, essential for understanding bimanual skill acquisition at the brain level [Gerloff].

Despite promising results, these neuroimaging approaches are not yet easily deployable due to their reliance on extensive preprocessing to extract skill-related neural biomarkers. The primary roadblock currently preventing an end-to-end solution is the substantial amount of preprocessing involved for data refinement and extraction of biomarkers. Specifically, fNIRS preprocessing typically involves the removal of motion artifacts and filtering of non-evoked and systemic signals associated with neurovascular coupling [Gao 2020] (Figure 1). This process begins with correcting the stimulus locations and pruning channels with low signal-to-noise ratios. Signals from the optodes are then converted to optical density (OD) for group-level analysis. At this stage, preprocessing focuses primarily on removing non-systemic noise and some systemic physiological noise. Non-systemic noise removal includes eliminating changes due to head movements, clenching, and instrument noise, while systemic noise removal targets physiological signals such as heartbeat, cardiac pulsations, respiration, and blood pressure oscillations [Yucel 2021]. Band-pass filtering removes most signals that fall outside the range of interest in fNIRS analysis, including signal drift. For further refinement, advanced methods like wavelet transform, empirical mode decomposition, and independent component analysis are often employed [Eastmond].

The cleaned signals are then used to estimate the concentration changes in oxyhemoglobin (HbO) and deoxyhemoglobin (HbR) using the modified Beer-Lambert law [Koscis 2006]. For subject-specific analysis, these signals are baseline corrected using signals prior to an experimental condition. Recent literature suggests using short separation (SS) signals to remove subject-specific confounding signals, applying generalized linear model (GLM) [von Luhmann 2020] for better generality of the observed response. on a trial-by-trial basis to create independent and identically distributed (iid) samples for assessing bimanual motor skills. Trial-wise signals are either averaged or feature-engineered (*e.g.* extracting statistical moments) [Gao 2021]. Some event-related study designs avoid segmentation altogether, instead directly extracting a hemodynamic response function (HRF) to model the response. This is done by convolving a stick function corresponding to the onset of individual trials. Other studies have used a feature extraction model, whereby trial-wise data are first transformed to spectrograms and Gramian angular fields and then used to extract the biomarkers/features using a multi-layer perceptron or convolutional neural network (CNN), though skipping some earlier preprocessing steps [Eastmond, 2022].

Traditional models for bimanual motor skill assessment often rely on these specialized preprocessing pipelines, which require extensive time and domain expertise. These models frequently struggle to generalize to new skills that introduce variables such as unaddressed motions, different tools, and subject-specific noise. This limitation stems from the fact that preprocessing steps are tailored specifically to the training data, leading to overfitting and making the models overly dependent on this data. Moreover, many existing models lack rigorous cross-validation and the ability to handle real-world variabilities, such as varying trial lengths and subjective noise, which are crucial for enhancing generalizability and robustness in realistic clinical settings.

To overcome these challenges, we propose an end-to-end solution based on deep learning that can directly process motor task-related physiological data without requiring domain-specific adjustments to the raw data stream. Our aim is to design and train a compact, low-parameter model that can generalize to unseen data from related procedures. While recurrent networks, *e.g.*, long short-term memory (LSTM), are used for data with variable length sequences [Cho, Rahul], we use a convolutional neural network (CNN) architecture due to its speed and versatility [Ronneberger, Ye]. The model extracts a low-dimensional representation or "context" of neuronal activation sequences using a convolutional encoder-decoder architecture. In the process, it transforms the raw brain imaging signals into a fully processed from, free from extracerebral signals, without requiring any intermediate signal processing steps. This extracted context of neuronal activation sequences is then used for the objective assessment of bimanual motor skills. Figure 1 illustrates a comparison of our approach with the existing paradigm for bimanual motor skill assessment in medicine.

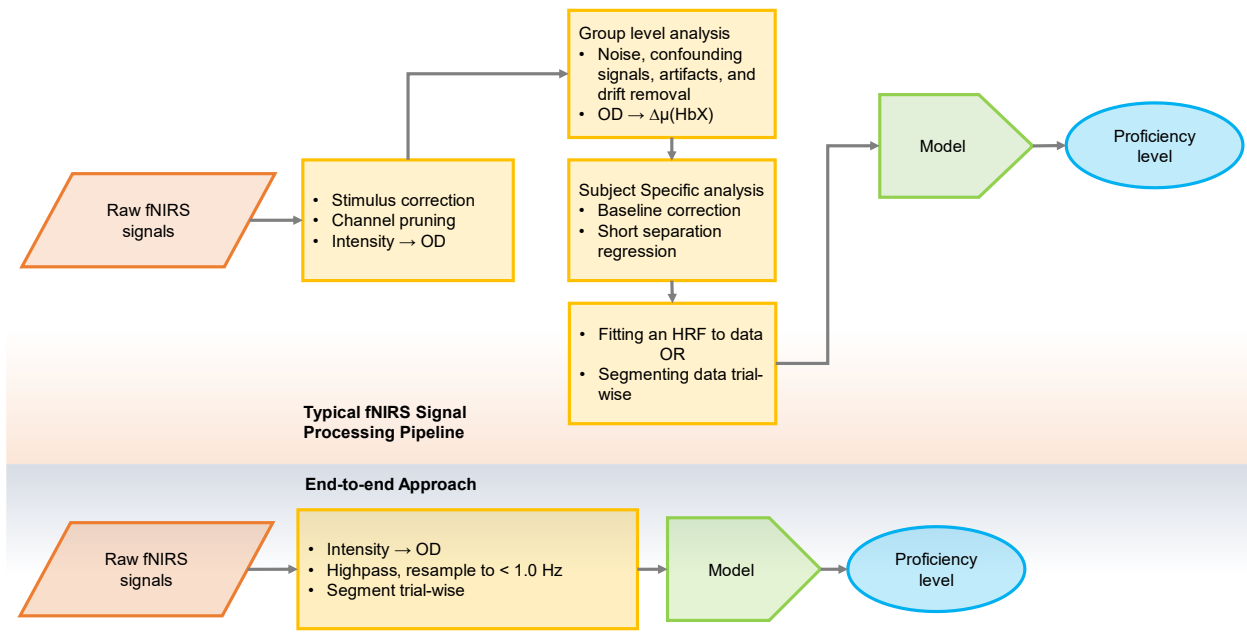

**Figure 1.** Comparison of (**TOP**) traditional fNIRS processing pipeline for bimanual skill assessment with (**BOTTOM**) our proposed end-to-end approach which directly uses raw fNIRS signals.

The remainder of this paper is organized as follows: the Methods section outlines the experimental design, and fNIRS data-acquisition and pre-processing steps followed by a detailed description of the end-to-end deep learning model including network training and validation steps. We focus our study on two laparoscopic surgical tasks, suturing and pattern cutting, part of the Fundamentals of Laparoscopic Surgery (FLS) program for board certification in general surgery, in addition to an emergency medicine endotracheal intubation (ETI) task. The Results section demonstrates the efficacy of our model in assessing skills from raw fNIRS signals, and the Discussion section analyzes its ability to distill the neural correlates associated with surgical and emergency medicine expertise. The paper concludes with future directions for this approach.

## 2. Methods

The end-to-end deep learning model for the assessment of bimanual motor skills was developed using neuroimaging datasets collected from three studies in which subjects performed complex bimanual motor tasks related to medical practice. Below, we present the study design, description of the datasets, data acquisition and processing, and the details of the network architecture, training and validation steps.

### 2.1. Study design

Human subject studies were conducted to collect brain imaging datasets for assessing bimanual motor skills in three tasks: Fundamentals of Laparoscopic Surgery (FLS) pattern cutting, FLS suturing with intracorporeal knot tying, and endotracheal intubation (ETI). The studies were approved by the University at Buffalo Institutional Review Board (IRB) and the US Army Human Research Protection Office (HRPO). All participants provided informed consent after a briefing on the experimental protocol. Neuronal activations were measured via fNIRS during task performance to identify neural correlates of bimanual motor skill proficiency. The study designs are shown in the block diagram in Figure 2 , and details of the four datasets used are summarized in Table 1.

**Table 1.** Experiments and Demographic Details.

| Study Arm and Task | Participants and Details | Training Protocol and Retention |
|---|---|---|
| **FLS Suturing: Learning Curve and Retention**<br>*Performance assessed using FLS scoring metrics [Fraser 2003]* | **Training group:** 28 students<br>Mean age: 23.57 years (SD 4.38)<br>11 males, 17 females<br>26 right-handed<br>**Control group**: 27 students<br>Mean age: 21.5 years (SD 2.5)<br>10 males, 17 females<br>24 right-handed | - 15-day training<br>- Reps per session: Days 1–5: 3, Days 6–10: 5, Days 11–15: 7<br>- Rest between attempts: 2 min<br>- Max duration per rep: 10 min<br>- Retention test after 4-week washout |
| **FLS Suturing: Expert-Novice Comparison**<br>*Performance assessed using FLS scoring metrics; study conducted separately from longitudinal study* | **Novices**: 15 students (from initial participant pool)<br>**Experts**: 15 surgeons<br>Average experience 4.53 years (SD 3.29) in laparoscopic surgery | - Single session<br>- Reps per session: 3 to 5<br>- Rest between attempts: 2 min<br>- Max duration per rep: 10 min<br>- Retention test: N/A |
| **Endotracheal Intubation (ETI)**<br>*Performance assessed based on successful intubation within allotted time* | **Training group:** 20 students<br>**Control group**: 19 students | - 3-day training<br>- Total reps: 10 over 3 days<br>- Rest between attempts: 2 min<br>- Max duration per rep: 3 min<br>- Retention test after 8-week interval |
| **FLS Pattern Cutting: Learning Curve and Retention**<br>*Performance assessed using FLS scoring metrics*<br>*Notes*: From sham cohort; details in [Kamat 2023, Gao 2021] | **Training group:** 7 medical students<br>Mean age: 24 years (SD 0.82)<br>2 males, 5 females<br>All right-handed | - 12-day training<br>- Reps per session: Days 1–11: 10, Day 12: 5<br>- Rest between attempts: 1 min<br>- Max duration per rep: 5 min<br>- Retention test after 4-week washout |

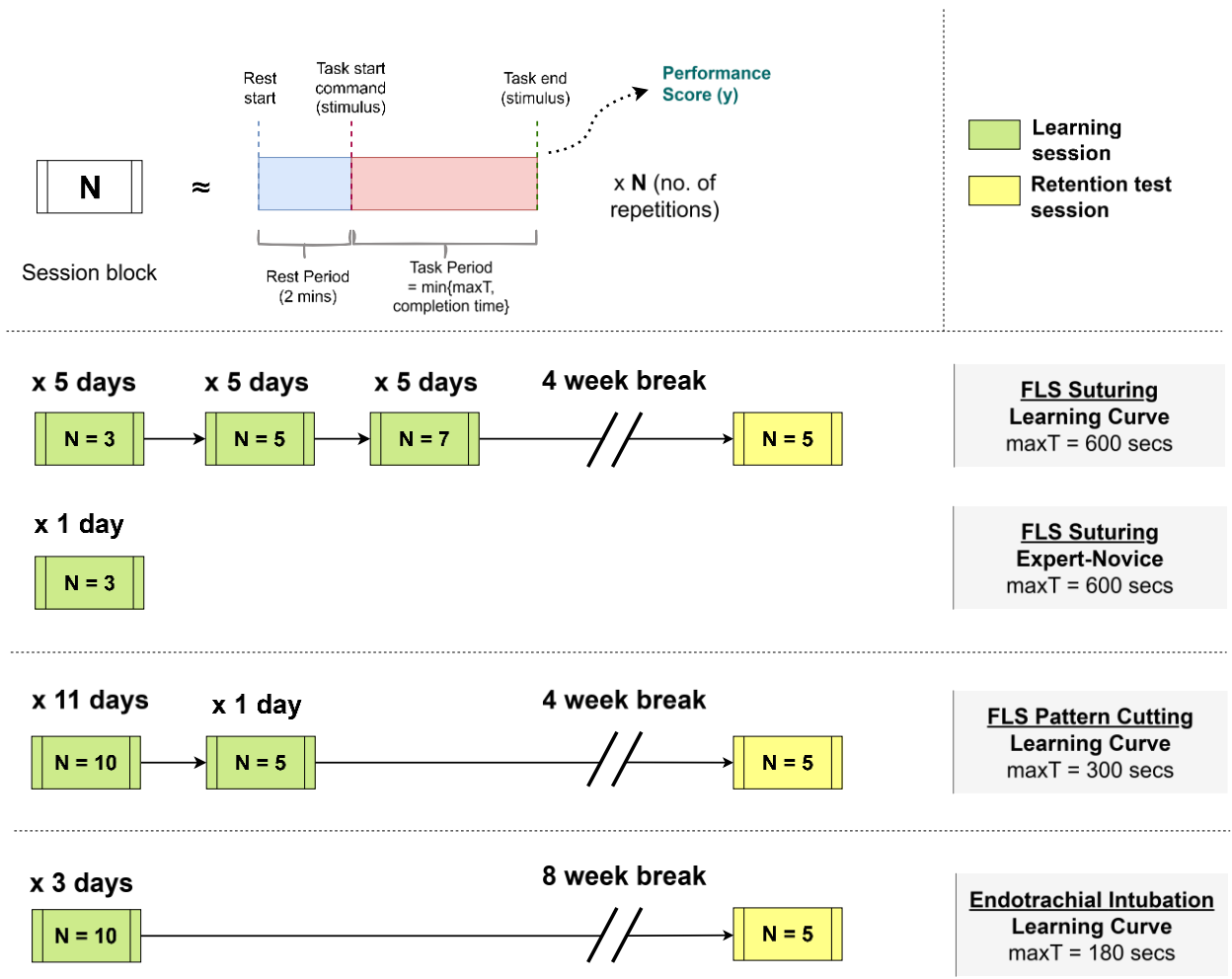

**Figure 2.** Experimental design for learning curve studies. Each session block (top row) includes N repetitions of the respective bimanual task. Each repetition consists a rest period followed by a task period, ending either at task completion or upon reaching maximum allowable time (maxT). Training for FLS suturing (middle row) spans 15 days, while ETI training (bottom row) is conducted over 3 days.

## 2.2. Data acquisition

The FLS suturing and ETI studies were conducted using the optode layout shown in Figure 3. The montage was configured based on the international 10-10 system [Jurcak]. This arrangement of optodes also allowed for concurrent EEG data acquisition whenever necessary. The headgear consisted of 16 sources and 15 detectors, placed 3 cm away from the sources. It also consisted of 8 short separation detectors placed 0.8 cm from sources. The equipment for data acquisition was a continuous-wave near-infrared spectrometer (NIRSport 2, NIRx Medical Technologies, LLC, NY, USA). This device sampled data at 5.0863 Hz and used infrared lights of wavelength 760 nm and 850 nm. The montage has coverage of 8 designated brain regions identified as left and right lateral sides of pre-frontal cortex (PFC), supplementary motor area (SMA), sensory-motor cortex (SMC) and parietal area (PAR). In total, there were 46 long-separation channels accompanied by 8 short-separation channels (see Figure 3).

The FLS pattern cutting studies were conducted using a custom montage based on a Monte-Carlo-optimized optode layout [Nemani 2014]. The montage (see Nemani 2014) has coverage of the brain regions identified as PFC, SMA and left and right motor cortices (MC).

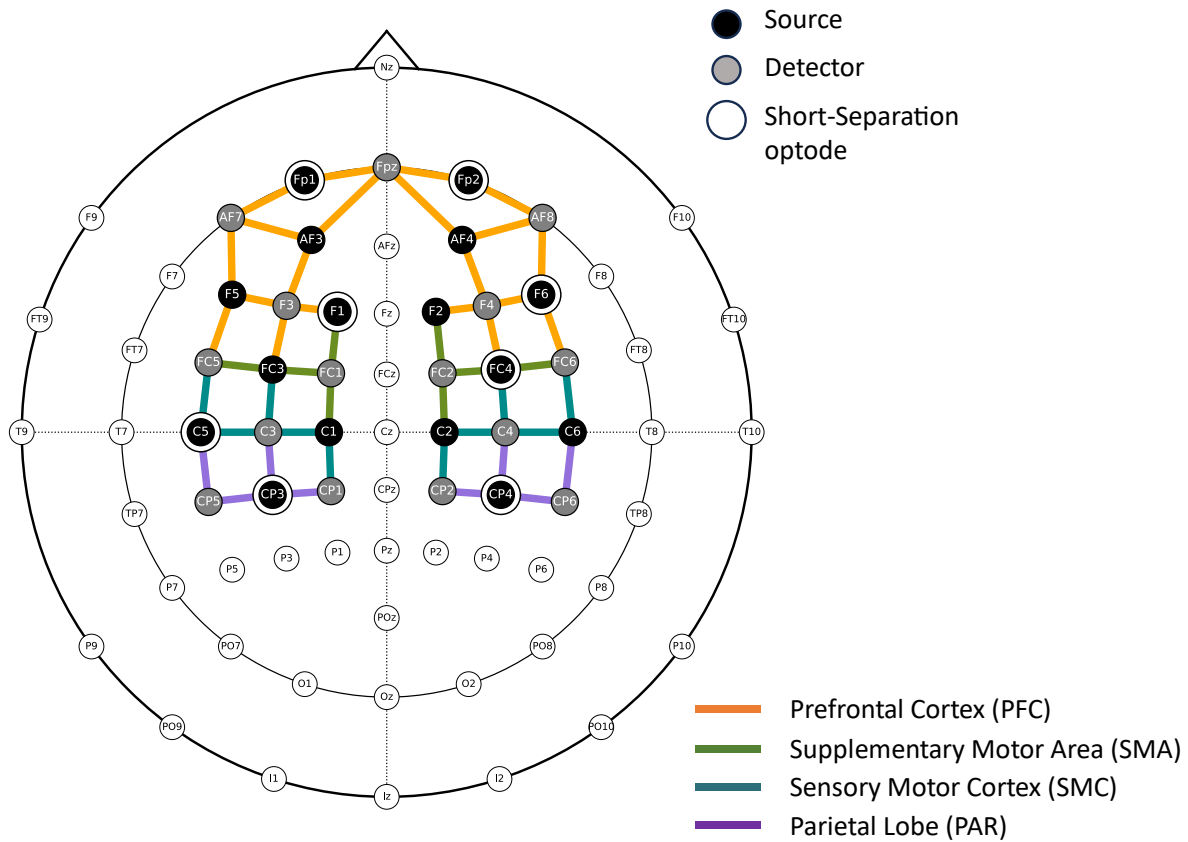

**Figure 3.** Montage for FLS-Suturing and ETI experiments with region-specific long-separation source-detector pairs color coded in legend.

The long-separation channels are used primarily to capture cortical activity, while the short-separation channels help regress out physiological signals from the measured data. Although it is widely acknowledged that long-separation channel measurements may contain confounding extracerebral signals [Yucel 2017, 2021], our focus is on cumulative markers of skill proficiency than isolating cognitive markers. An added benefit of using physiological data is the inclusion of markers related to non-technical skills, such as situational awareness and decision-making [Elek]. Furthermore, Generalized Linear Model (GLM)-based approaches [Cohen Adad, von Luhmann 2020], being parametric, are subject to variation across demographic studies. As our model is designed to avoid experiment-specific tuning, we implement a fixed processing pipeline across the three experiments using the same montage.

## 2.3. fNIRS signal processing

To develop the end-to-end pipeline, we need to implement the most generic processing that needs to be applied on the raw data extracted from the equipment. This begins by converting changes in light intensity to changes in optical density, followed by band-limiting the signals to the hemodynamic activity range. Given the slow nature of the hemodynamic response, applying a band-pass filters from 0.01 to 0.1 Hz removes most non-evoked signals [Scholkman 2014], isolating the relevant signal with minimal interference from non-task-related noise.

The second stage of preprocessing involves more advanced corrections, such as motion artifact removal, which is performed using spline polynomial fitting in Homer3 v1.33.0 [Huppert]. Changes in light intensity are then converted to oxy- and deoxyhemoglobin concentrations, using partial pathlength factors of (0.1, 0.1) [Cope 1988]. These concentrations represent chromophore activity within cortical areas, reflecting oxygen consumption linked to neural activity. The fully processed signals serve as ground truth for model optimization, allowing the model to learn the low-dimensional representation necessary for accurate signal transformation.

## 2.4. Datasets

A total of 56 subjects participated in the learning curve studies for three bimanual motor tasks. The control group for the FLS suturing task included 25 participants. Subjects who discontinued, as well as left-handed individuals and those with thick scalp hair that introduced excessive noise, were excluded. The trials exceeding pre-specified time-limits or obtaining zero scores for the FLS tasks were disqualified and discarded. Additionally, trials with signal lengths deviating significantly from the recorded acquisition times in a separate real-time log were identified as outliers and excluded. Trials shorter than 17 time points, due to downsampling, were also removed, as they were incompatible with the network's convergence. The number of successful and unsuccessful trials for the tasks after exclusion is presented in Table 2. For brevity, the dataset from the FLS suturing learning curve and retention studies will be referred to as FLS-S and FLS-SRet, respectively. Similarly, datasets for the FLS pattern cutting tasks are denoted as FLS-PC and FLS-PCRet, and for the ETI task as ETI and ETI-Ret, respectively.

**Table 2.** Dataset after exclusion.

| Dataset | Number of trials | | Subjects | Montage system | Distinct brain regions | Number of long channels |
|---|---|---|---|---|---|---|
| | Unsuccessful (+ve) | Successful (-ve) | | | | |
| FLS-S | 1207 | 236 | 21 (LC) + 25 (C) | 10-10 | 8 | 46 |
| FLS-SRet | 43 | 15 | 21 (LC) | 10-10 | 8 | 46 |
| ETI | 127 | 357 | 20 (LC) + 16 (C) | 10-10 | 8 | 46 |
| ETI-Ret | 4 | 31 | 12 (LC) | 10-10 | 8 | 46 |
| FLS-PC | 43 | 587 | 7 | Custom | 5 | 28 |
| FLS-PCRet | 0 | 19 | 7 | Custom | 5 | 28 |

LC: learning curve; C: Control.

## 2.5. End-to-end deep learning model

### 2.5.1. Network architecture

The deep learning model is based on a one-dimensional convolutional encoder-decoder neural network architecture [Ronneberger] that maps raw fNIRS signals to fully processed outputs. This mapping is analogous to processing raw neuronal activation sequences into a processed form devoid of motion artifacts and extracerebral noise. Through this process, the network learns a low-dimensional representation of the fully processed signal that encapsulates the context (or biomarker) that is sufficient to reconstruct the neuronal activation sequences. A unique aspect of the model is that it distills the context from subject-specific neuronal activation sequence that is often of variable length. The context is then utlized for the classification of bimanual motor skill level. The architecture is depicted in Figure 5.

The encoder consists of three encoding blocks with one-dimensional convolutional layers. The 1D convolutional filters detect time-series patterns, such as periodic spikes, to generate a feature map, with subsequent layers capturing higher-level abstractions. The first convolutional layer uses a large kernel size of 11 and a stride of 1 in the temporal direction, allowing it to span the typical hemodynamic response function's peak, dip, and return to baseline over 15–20 seconds [Strangman]. Two additional convolutional layers with a kernel size of 3 and stride of 1 further refine the features. Each convolved feature map is activated using the SeLU function [Klambauer]. No pooling layers are includes, as subsampling is not necessary. The feature maps are causal-padded to preserve dimension up to the bottleneck layer, after which the padded is trimmed. Squeeze-Excitation (SE) blocks [Hu 2018] assign importance to each convolution filter, with activation provided by either ReLU (default) or GELU functions [Hendrycks] to enhance feature selection.

The decoder consists of three convolution-transpose layers with kernel size 3, 3 and 11, all with stride 1 to match the output time-series size. Each layer uses the SeLU activation function. The reconstructs the processed ($\Delta HbO$/$\Delta HbR$) data from the refined feature map. To enhance learning and increase robustness of filters, one-sixth of the filters are randomly dropped during each training iteration. The entire model has approximately 7000 parameters. Global average pooling is applied to the feature map to generate a contextual vector of size 24, which is used to train the classifier.

The classifier consists of the pretrained encoder, followed by a global average pooling layer and a one-hidden-layer neural network (1NN) that outputs softmaxed logits for each predicted class. The 1NN establishes a decision boundary to distinguish between successful and unsuccessful classes. It contains 120 nodes in the first hidden layer, with L1-L2 regularization and a 0.5 dropout rate to mitigate overfitting.

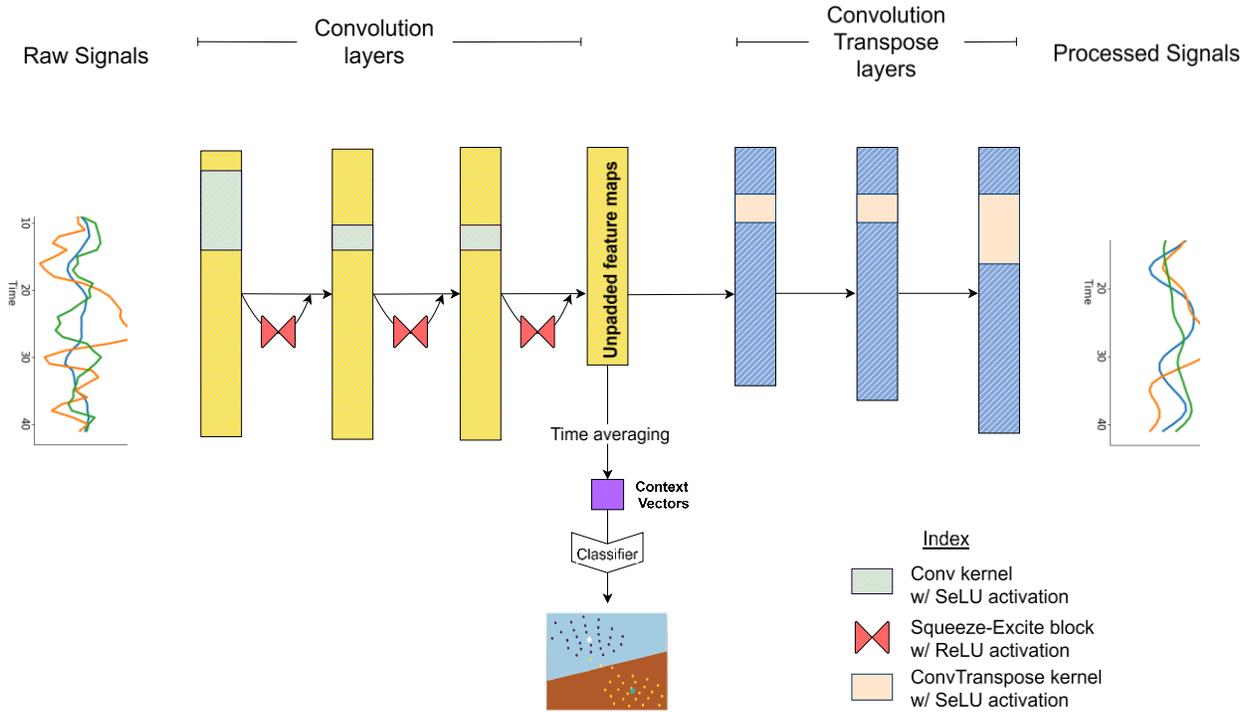

**Figure 5.** Encoder-decoder 1DCNN architecture for representation learning.

### 2.5.2. Training and validation

The 1DCNN encoder-decoder network is trained in a self-supervised fashion to produce fully processed signal from the raw fNIRS measurements. The 1NN classifier is then trained on contextual features (or biomarkers) extracted from the neuronal activation sequences associated with successful or unsuccessful surgical (FLS) or emergency medicine (ETI) procedure, as measured by the well-established metrics.

The encoder-decoder network's input consists of band-limited optical density data belonging to two wavelengths of near infrared light, from each fNIRS channel. Each input sample has dimension of $(\tau \times 2n_{ch})$, where $\tau$ is the trial duration for each subject, and $n_{ch}$ is the number of long separation channels covering a particular brain region. Given the variability in $\tau$ across trials, all examples in a batch are padded to match the longest sequence in that batch, with end-padded values masked during loss calculation. Mean-squared-error (MSE) is computed only on the first 25 timepoints of each example. The raw neuronal activation sequences are normalized within a 0-1 range for each channel independently before training, and the same normalization scale is applied to test data. During pretraining, output signals are individually normalized based on channel-wise statistics of the ground truth signals, which improves reconstruction. After training, the pretrained encoder with frozen weights is combined with the classifier, yielding a dataset of $N_{samples}$ x 24, which serves as input to the classifier. The training and inference processes are illustrated in Figure 6.

The encoder-decoder network is trained to minimize the MSE using the Adam optimizer with stochastic gradient descent [Kingma and Ba] with a cyclical learning rate [Smith] between $5 \times 10^{-4}$ and $10^{-2}$ for consistent convergence. The network is trained for a maximum of 6000 epochs using a batch size of 32, and the weights are saved at checkpoints when the reconstruction loss is minimum. Training is stopped early if loss does not decrease for more than 250 epochs. The classifier is trained using the same learning rate schedule and batch size of 32 for 1500 epochs per cross-validation fold. In each fold, the validation set is either a subset of the full dataset (in k-fold) or data from a subject excluded from the training set (Leave-One-Subject-Out). Metrics are computed for each fold to report the mean and standard deviation. The model is implemented using Keras library [*Keras* ] and trained on TensorFlow's computational graph [Abadi ].

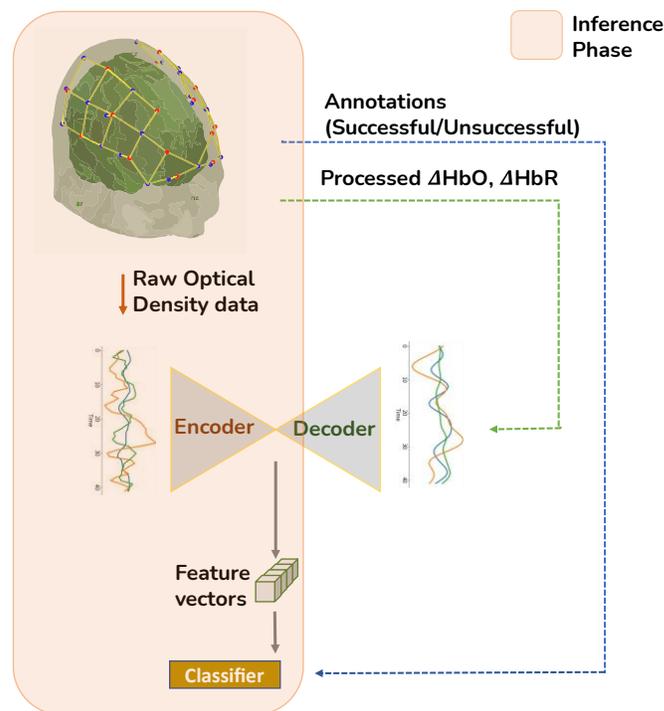

**Figure 6.** Schematic of the model training, with inference processes highlighted. During training, raw optical density data is input to the encoder-decoder IDCNN to reconstruct a clean chromophore concentration change signal. Bottleneck activations are time averaged to generate a feature vector for classification into two groups. The shaded inference phase indicates the use of the trained encoder and classifier on raw skill-retention datasets to obtain output labels.

The end-to-end model is cross validated on the training set and tested for generalization on the retention dataset with a similar data distribution. We perform a 15-fold cross-validation (CV) to evaluate model performance on the training set, reporting central tendency measures for each fold. Additionally, we test generalization directly on the skill retention datasets for each fold. To further assess the model's ability to generalize to new subjects, we conduct Leave-One-Subject-Out (LOSO) CV[Esterman]. In this approach, the model is trained on the dataset from all but one subject, which is used for testing. This process is repeated until the model has been independently assessed on all the subjects.

### 2.5.3. Model evaluation metrics

The performance of the end-to-end model was evaluated using metrices such as Matthew's correlation coefficient (MCC), F1 score, sensitivity, specificity, and accuracy, derived from the confusion matrix [Chicco]. For the bimanual motor skill classification, trials labeled as successful based on predefined scoring metrics were treated as negative samples. Performance metrics were averaged across all cross-validation folds and reported in the results. To address class imbalance, a higher weightage (70%) was assigned to the loss of the minority class during training.

The 1NN classifier generates predictions as confidence scores for each output class, which reflects the model's confidence in its prediction. These confidence scores also facilitate the evaluation of the classifier's reliability through the area under the Precision-Recall (PR) curve and the Receiver-Operating Characteristics (ROC) curve. A fixed threshold of 0.5 was applied across all datasets to classify predictions as either 0 (negative) or 1 (positive).

### 3. Results

The effectiveness of the end-to-end deep learning model was evaluated for its ability to assess bimanual motor skill learning in surgery and emergency medicine tasks. To benchmark its performance, a comparative study was performed using a baseline model with the same network architecture as the end-to-end model. Unlike the end-to-end approach, the baseline model processes fully processed fNIRS signals to reconstruct themselves, with the extracted context used to classify skills levels. This comparison aimed to test the hypothesis that bimanual motor skill assessment using raw brain imaging data can achieve performance metrics comparable to or better than that those obtained from a baseline model relying on fully preprocessed signals.

The loss curves and reconstruction performance of the encoder-decoder architecture are provided in the supplementary material.

### 3.1. FLS suturing task

The end-to-end model was evaluated on the FLS-S dataset with the unprocessed input fNIRS signals to differentiate between successful and unsuccessful FLS suturing tasks in a learning curve study. The performance metrics obtained from 15-fold cross validation are presented in Table 3. The model demonstrated greater than 87% accuracy and sensitivity across all cortical regions, indicating a strong ability to classify unsuccessful tasks correctly. Specifically, at most 151 out of 1207 unsuccessful tasks were incorrectly classified as successful, while at most 48 out of 236 successful trials were misclassified as unsuccessful.

The area under the curve (AUC) for both ROC and PR curves (Figure 8) highlights the strong discriminative power of brain activations across all regions in distinguishing between successful and unsuccessful FLS suturing tasks. Specifically, the ROC AUC values range from 0.93 to 0.94 across brain regions, with RPFC showing the highest mean value ($0.94 \pm 0.02$), indicating excellent overall classification performance. Similarly, the PR AUC values are consistently high (0.99 for all regions with negligible variability), reflecting the model's ability to maintain precision even under class imbalance conditions.

These findings underscore the capability of the end-to-end model to effectively process raw fNIRS signals and extract meaningful neural correlates for skill assessment. The narrow confidence intervals in both plots suggest the model's reliability across validation sets, further solidifying its robustness for real-world applications in assessing bimanual motor skill learning.

**Table 3.** End-to-end model mean performance metrics obtained from 15-fold cross-validation.

| Brain Region | MCC | F1-score | Sensitivity | Specificity | Accuracy | ROC-AUC | PR-AUC |
| --- | --- | --- | --- | --- | --- | --- | --- |
| LPFC | 63.8% | 92.5% | 88.9% | 83.4% | 88.0% | 0.933 | 0.988 |
| RPFC | 63.8% | 93.1% | 90.5% | 79.9% | 88.8% | 0.935 | 0.989 |
| LSMA | 64.5% | 92.8% | 89.4% | 83.0% | 88.4% | 0.932 | 0.988 |
| RSMA | 62.8% | 92.5% | 89.3% | 80.9% | 87.9% | 0.934 | 0.988 |
| LSMC | 65.7% | 92.9% | 89.3% | 84.7% | 88.6% | 0.934 | 0.988 |
| RSMC | 62.8% | 93.0% | 90.7% | 77.6% | 88.6% | 0.932 | 0.988 |
| LPAR | 60.3% | 92.6% | 90.5% | 75.1% | 87.9% | 0.929 | 0.987 |
| RPAR | 60.2% | 92.6% | 90.6% | 74.2% | 87.9% | 0.93 | 0.988 |

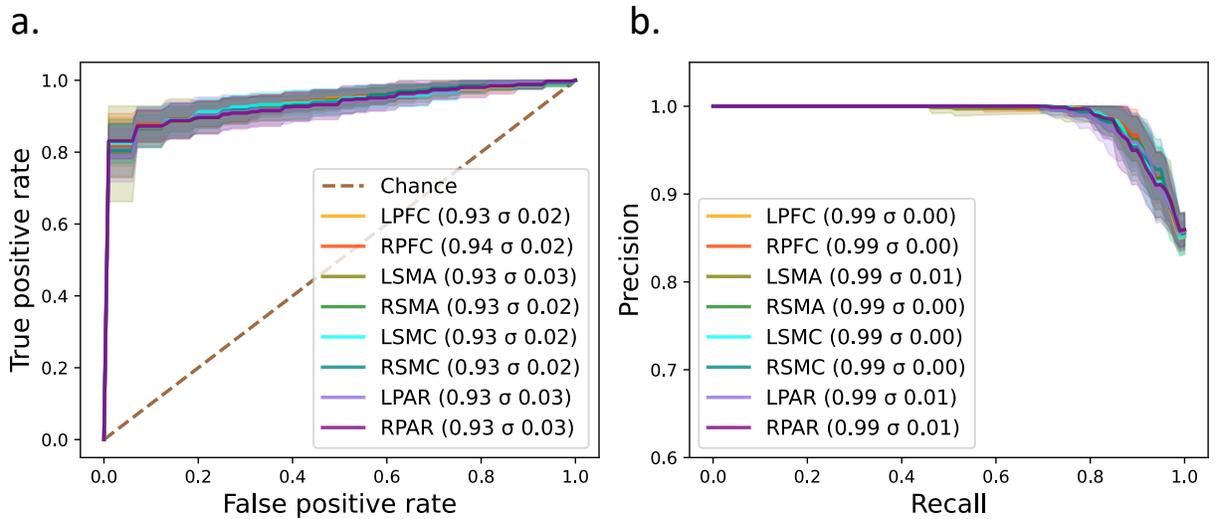

**Figure 8:** (a) receiving operator characteristics (ROC) and (b) precision-recall (PR) curves obtained from 15-fold cross validation performed on the FLS-S training dataset for each brain region using the end-to-end model. Shaded regions denote confidence intervals. AUC values are reported with standard deviation.

The generalizability of the model was evaluated on the FLS-Ret dataset, with performance metrics presented in Table 4. The end-to-end model demonstrated strong generalization across all cortical regions, effectively differentiating skill retention levels. Notably, RPFC activations yielded the highest AUC values (0.99 ± 0.00) in both the ROC and PR curves (Figure 9), underscoring their strong discriminative capability. Similarly, LSMC activations also exhibited excellent performance, with high ROC AUC (0.97 ± 0.00) and PR AUC (0.99 ± 0.00) values, confirming their reliability for skill classification.

Although LPFC activations had slightly lower ROC AUC (0.92 ± 0.01), their PR AUC remained high (0.97 ± 0.01), suggesting robust precision even with a slight trade-off in recall. The uniformity of performance across regions, with all AUC values above 0.95, highlights the model's strong generalizability and consistency. These findings validate the model's capability to generalize well to unseen retention data, further supported by the narrow confidence intervals observed in the plots.

Table 4. End-to-end model mean performance metrics for the test (retention) dataset.

| Brain Region | MCC | F1-score | Sensitivity | Specificity | Accuracy | ROC-AUC | PR-AUC |
|---|---|---|---|---|---|---|---|
| LPFC | 65.8% | 91.9% | 96.3% | 62.2% | 87.5% | 0.923 | 0.971 |
| RPFC | 79.0% | 94.3% | 94.0% | 84.9% | 91.6% | 0.964 | 0.989 |
| LSMA | 74.8% | 93.7% | 95.2% | 77.3% | 90.6% | 0.951 | 0.985 |
| RSMA | 76.6% | 92.7% | 94.1% | 78.2% | 90.0% | 0.957 | 0.985 |
| LSMC | 81.6% | 95.5% | 97.7% | 80.0% | 93.1% | 0.97 | 0.991 |
| RSMC | 78.4% | 93.6% | 92.2% | 87.1% | 90.9% | 0.974 | 0.993 |
| LPAR | 80.1% | 95.1% | 96.9% | 80.0% | 92.5% | 0.965 | 0.99 |
| RPAR | 79.1% | 94.5% | 95.0% | 82.7% | 91.8% | 0.961 | 0.987 |

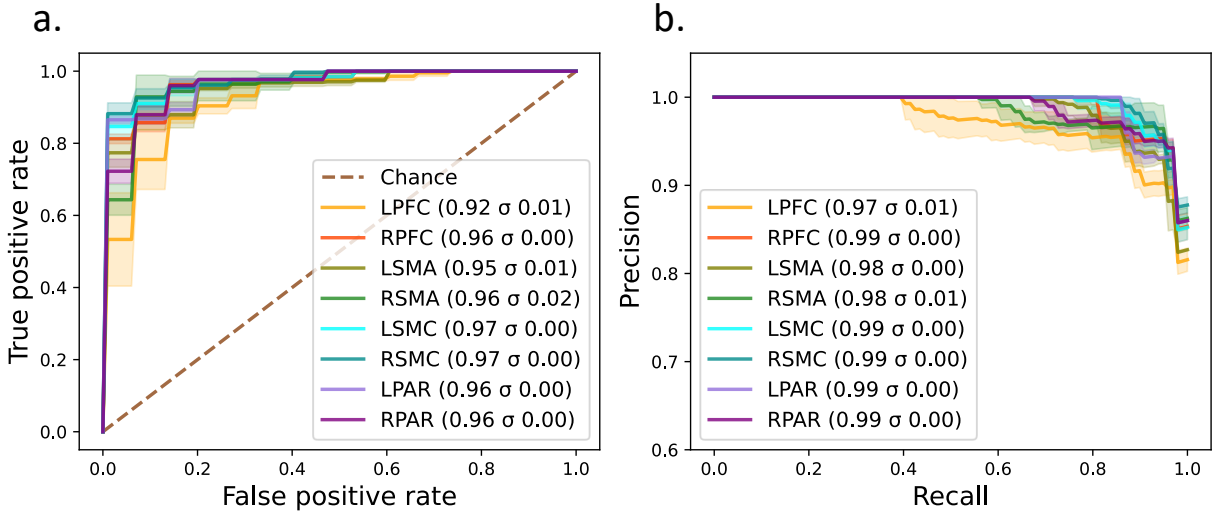

**Figure 9:** (a) ROC and (b) PR curves obtained from inference performed on the FLS-Ret dataset for each brain region during 15-fold cross validation using the end-to-end model. Shaded regions denote the confidence intervals. Mean AUC values are reported with standard deviations.

A similar study was conducted using the baseline model trained and tested on fully processed fNIRS signals. Performance metrics for both the 15-fold cross-validation on the training set (FLS-S) and generalizability test conducted on the retention dataset (FLS-Ret) are provided in Tables 5 and 6, respectively, with the ROC and PR curves are shown in Figure 10. The results indicate that the end-to-end model outperformed the baseline model on raw fNIRS signals across most cortical regions.

Notably, RPFC activations exhibited reduced discriminative power with the baseline model, while LPFC showed relatively better generalization. This indicates that removing physiological noise from LPFC activations may enhance performance, particularly for skill retention testing, if LPFC is a region of interest. Conversely, RSMA performance decreased significantly with the baseline model, suggesting that preprocessing might obscure critical features in this region, reducing classification accuracy. While LPFC activations showed improved generalization with the baseline model, the specificity of other brain regions was notably reduced compared to the end-to-end approach, highlighting the robustness of the latter in preserving relevant signal characteristics for accurate skill classification.

Furthermore, while LPFC activations demonstrated improved generalization ability under the baseline model, other regions showed reduced specificity compared to the end-to-end approach. The narrower confidence intervals and higher AUCs observed in the end-to-end model indicate its robustness in preserving signal characteristics critical for accurate skill classification. Overall, these findings emphasize the advantage of using raw fNIRS data with the end-to-end model, particularly for regions like RPFC and RSMA, where preprocessing may hinder performance.

Table 5. Baseline model mean performance metrics obtained from 15-fold cross-validation.

| Brain Region | MCC | F1-score | Sensitivity | Specificity | Accuracy | ROC-AUC | PR-AUC |
|---|---|---|---|---|---|---|---|
| LPFC | 61.3% | 92.7% | 90.5% | 76.2% | 88.1% | 0.925 | 0.986 |
| RPFC | 54.5% | 91.7% | 90.0% | 68.2% | 86.4% | 0.916 | 0.984 |
| LSMA | 56.9% | 91.7% | 89.1% | 73.8% | 86.6% | 0.916 | 0.985 |
| RSMA | 56.9% | 92.1% | 90.1% | 71.2% | 87.0% | 0.915 | 0.984 |
| LSMC | 60.0% | 92.5% | 90.4% | 74.5% | 87.8% | 0.924 | 0.986 |
| RSMC | 59.1% | 92.1% | 89.5% | 75.4% | 87.2% | 0.921 | 0.985 |
| LPAR | 58.6% | 92.4% | 90.6% | 72.1% | 87.6% | 0.918 | 0.984 |
| RPAR | 60.2% | 92.5% | 90.4% | 74.6% | 87.8% | 0.917 | 0.984 |

Table 6. Baseline model mean performance metrics for the test (retention) dataset.

| Brain Region | MCC | F1-score | Sensitivity | Specificity | Accuracy | ROC-AUC | PR-AUC |
|---|---|---|---|---|---|---|---|
| LPFC | 83.1% | 95.6% | 96.0% | 86.2% | 93.4% | 0.925 | 0.961 |
| RPFC | 77.0% | 93.8% | 93.5% | 83.6% | 90.9% | 0.955 | 0.985 |
| LSMA | 80.6% | 95.2% | 97.4% | 79.6% | 92.8% | 0.965 | 0.988 |
| RSMA | 53.6% | 87.9% | 88.4% | 64.0% | 82.1% | 0.872 | 0.952 |
| LSMC | 77.1% | 94.3% | 95.7% | 79.1% | 91.4% | 0.964 | 0.989 |
| RSMC | 68.5% | 91.9% | 92.1% | 76.0% | 87.9% | 0.925 | 0.974 |
| LPAR | 65.7% | 92.0% | 96.1% | 63.1% | 87.6% | 0.948 | 0.983 |
| RPAR | 79.8% | 95.0% | 96.9% | 79.6% | 92.4% | 0.951 | 0.983 |

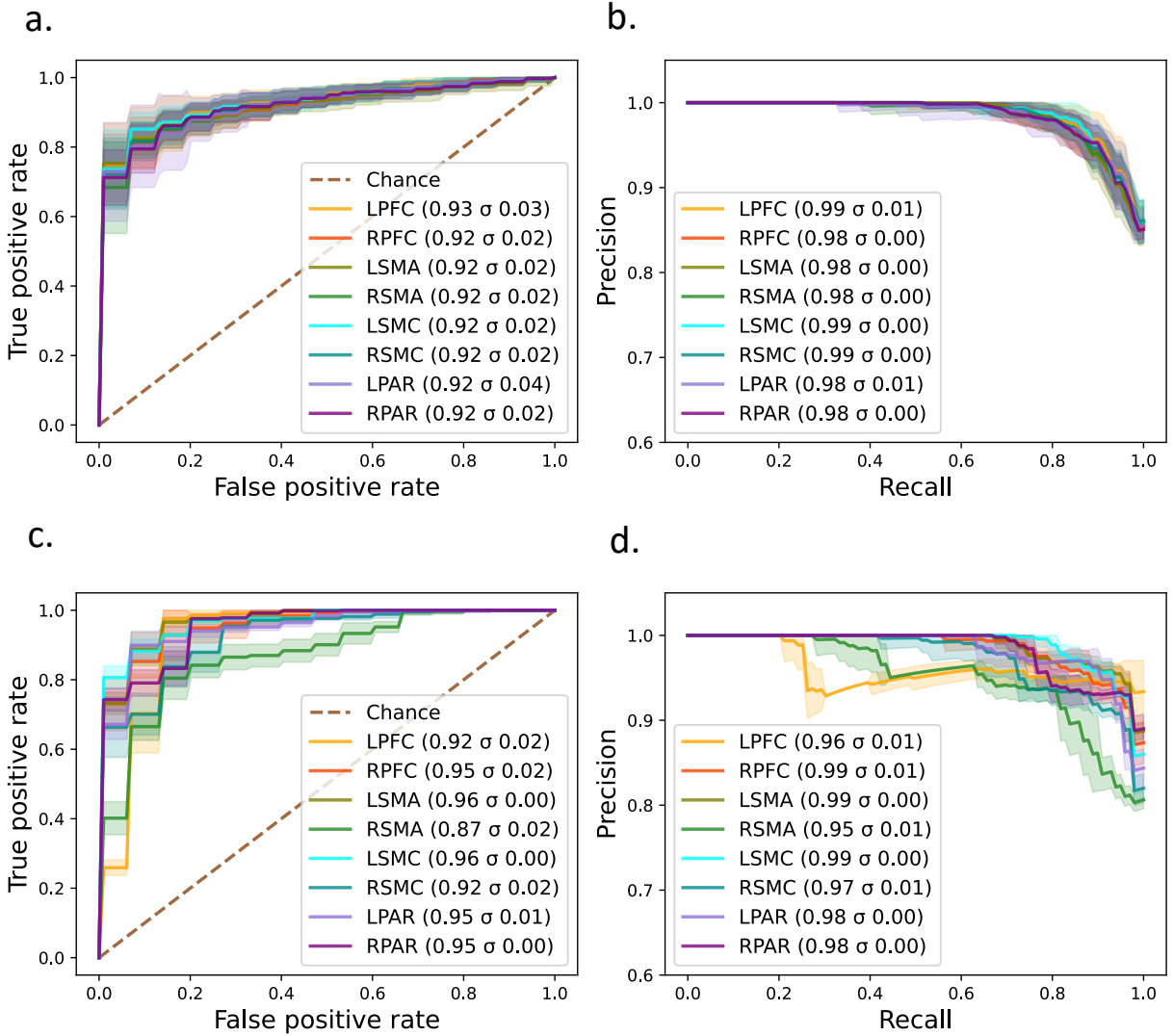

**Figure 10:** (a) ROC and (b) PR curves obtained from 15-fold cross-validation performed on the FLS-S training set for each brain regions using baseline model. (c) ROC and (d) PR curves from a similar study performed on the FLS-Ret test (retention) set for each brain regions. Shaded regions denote the confidence regions. Mean AUC values are reported with standard deviation.

### 3.2. ETI emergency medicine task

The ETI dataset yielded strong performance across all brain regions in the training set, with AUC values for the ROC curves consistently approaching 0.98 or higher, as shown in Figure 10a. The high specificity values, nearly reaching 100% (Table 7), underscore the model's capability to accurately detect false positives. The PR curves (Figure 10b) further confirm the model's reliable performance with minimal variance, particularly in regions like LPFC and RPFC.

However, since nearly all subjects in the retention test successfully completed the emergency medicine task, the dataset contains predominantly negative examples. Consequently, only accuracy is reported for the retention set to assess the model's generalizability. Accuracy remains high across all brain regions for both the training and retention datasets, with consistent metrics indicating the robustness of the end-to-end model for this task. Despite the insufficient number of samples available in the minority class to generate reliable PR and ROC curves, sensitivity and specificity are reported in the supplementary section (Table S10), both of which exceed 90%. These findings emphasize the model's ability to generalize effectively despite imbalanced class distributions in the retention set.

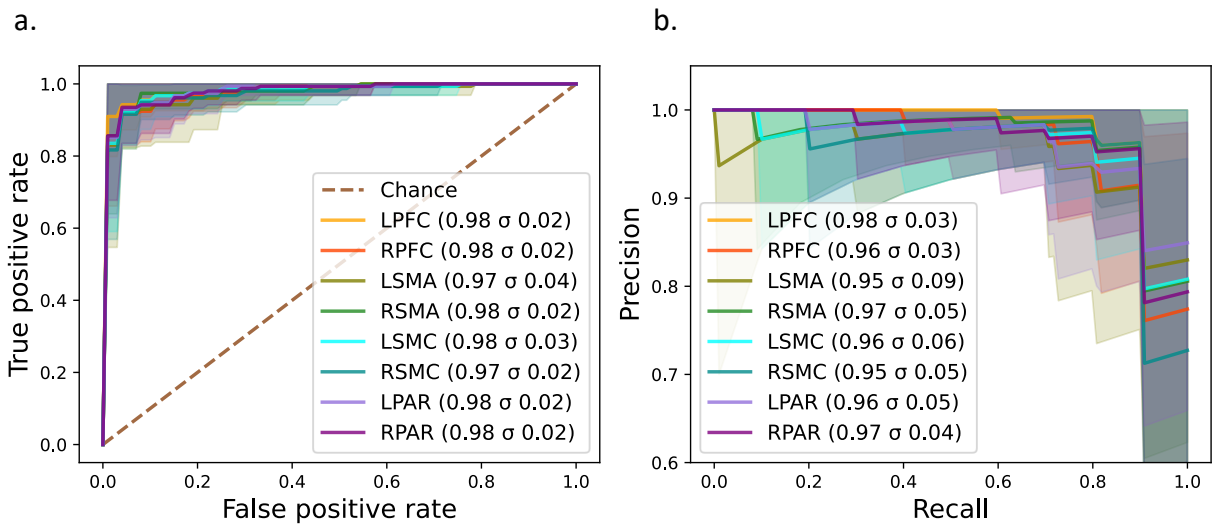

**Figure 10:** (a) ROC and (b) PR curves obtained from 15-fold cross-validation performed on the ETI training set for each brain region using the end-to-end model. Shaded regions represent confidence intervals, with the mean AUC values and standard deviations reported.

Table 7. End-to-end model mean performance metrics obtained from 15-fold cross-validation performed on the ETI dataset.

| Brain Regions | 15-fold Cross-validation Results ||||||| Test Set Results |
|---|---|---|---|---|---|---|---|---|
| | MCC | F1-score | Sensitivity | Specificity | Accuracy | ROC-AUC | PR-AUC | Accuracy |
| LPFC | 92.1% | 94.0% | 91.6% | 98.8% | 96.9% | 0.98 | 0.975 | 95.0% |
| RPFC | 92.7% | 94.2% | 90.6% | 99.5% | 97.0% | 0.975 | 0.961 | 93.3% |
| LSMA | 91.5% | 93.5% | 91.1% | 98.8% | 96.7% | 0.973 | 0.949 | 93.7% |
| RSMA | 92.1% | 93.7% | 89.5% | 99.5% | 96.9% | 0.979 | 0.965 | 92.8% |
| LSMC | 91.5% | 93.3% | 89.6% | 99.3% | 96.7% | 0.977 | 0.96 | 91.2% |
| RSMC | 91.1% | 93.2% | 90.4% | 98.8% | 96.5% | 0.972 | 0.954 | 91.4% |
| LPAR | 90.8% | 92.7% | 89.9% | 98.8% | 96.3% | 0.978 | 0.964 | 92.6% |
| RPAR | 91.6% | 93.3% | 89.6% | 99.3% | 96.7% | 0.978 | 0.966 | 94.7% |

When compared to the baseline model results (Figure 11), the end-to-end model demonstrates consistently better performance across all brain regions, particularly in the pre-frontal regions such as LPFC and RPFC, as evidenced by the higher AUC values for both the ROC and PR curves (Figure 10). Notably, the end-to-end model exhibits less variability in precision-recall performance compared to the baseline model, particularly in regions like LPAR and RPAR. This indicates its robustness in handling signal variability and extracting meaningful features directly from raw fNIRS data. Detailed performance metrics for the baseline model are provided in the supplementary section.

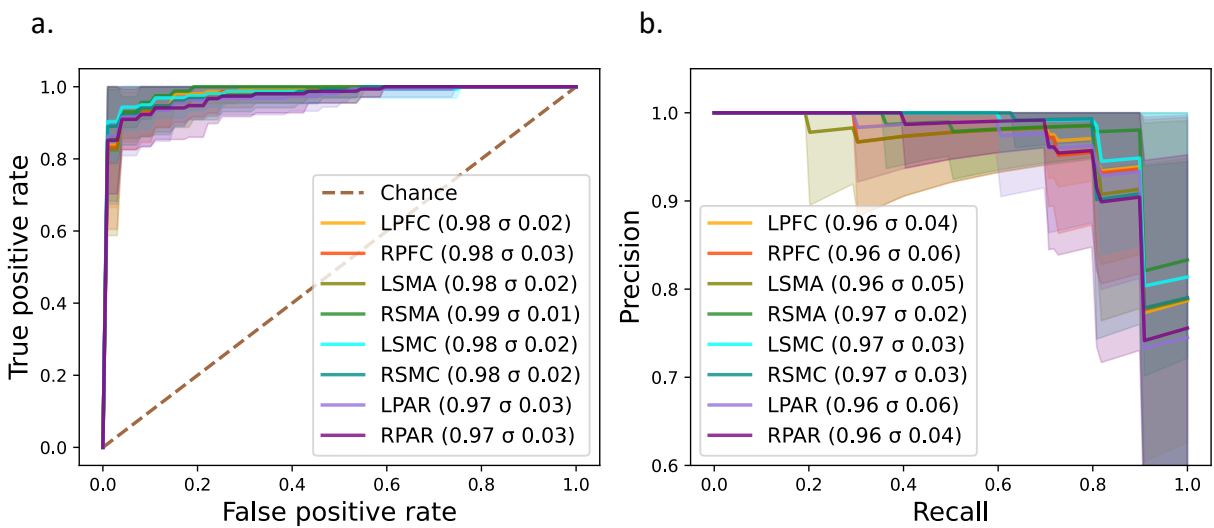

**Figure 11:** (a) ROC and (b) PR curves from 15-fold cross-validation conducted on the ETI training set for each brain region using the baseline model. Shaded regions represent confidence intervals, with mean AUC values and standard deviations reported.

### 3.3. FLS pattern cutting task

For the FLS-PC dataset, RPFC activations yielded the best results for both training and test (retention) sets, despite a 1:14 imbalance ratio in the data as shown in Table 8. The PR curves in Figure 12 reveal that LPFC activations shows higher variance and lower AUC compared to the other regions. While LPFC accuracy on the training set exceeded 98% its sensitivity was lower, likely due to the overrepresentation of negative (successful) examples. This suggests that LPFC activations are more prone to false negatives in the presence of class imbalance, while RPFC activations remain robust. Notably, all participants in the study were right-handed, which typically results in more prominent LPFC activations, making this differential effect noteworthy. Curves for the test sets were not generated due to insufficient samples in both binary classes.

**Table 8.** End-to-end model mean performance metrics obtained from 15-fold cross-validation performed on the FLS-PC dataset.

| Brain Regions | 15-fold Cross-validation Results | | | | | | | Test Set Results |
| --- | --- | --- | --- | --- | --- | --- | --- | --- |
| | MCC | F1-score | Sensitivity | Specificity | Accuracy | ROC-AUC | PR-AUC | Accuracy |
| **LPFC** | 85.3% | 86.2% | 87.8% | 98.8% | 98.1% | 0.99 | 0.92 | 94.7% |
| **RPFC** | 93.4% | 93.6% | 97.8% | 99.1% | 99.0% | 0.99 | 0.99 | 94.4% |
| **SMA** | 89.7% | 90.0% | 91.1% | 99.1% | 98.6% | 0.99 | 0.94 | 94.7% |
| **LMC** | 89.3% | 89.7% | 91.1% | 99.0% | 98.4% | 0.99 | 0.96 | 95.1% |
| **RMC** | 92.4% | 92.6% | 97.8% | 99.0% | 98.9% | 0.99 | 0.95 | 94.0% |

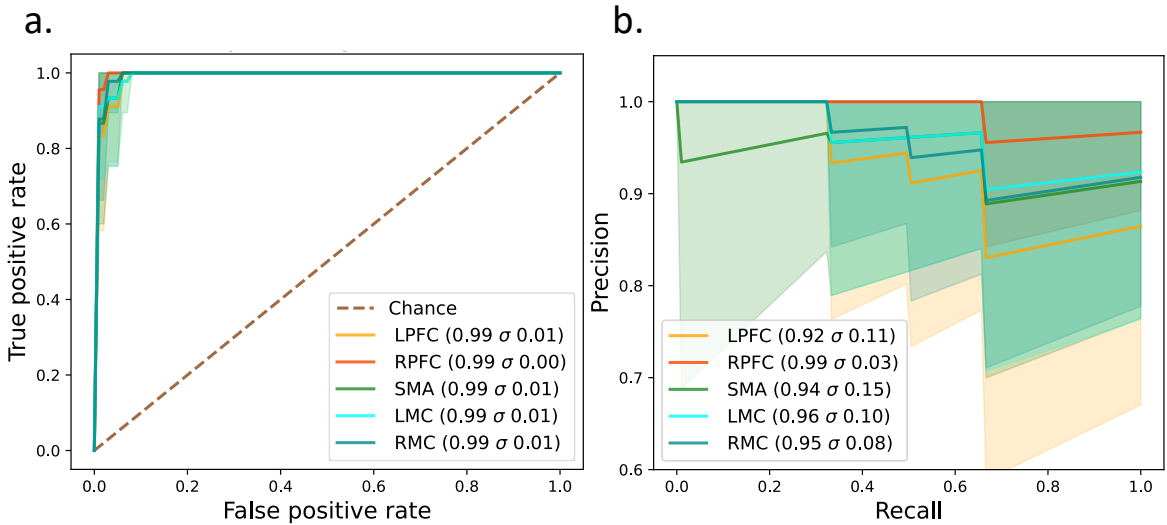

**Figure 12:** (a) ROC and (b) PR curves from 15-fold cross-validation performed on the FLS-PC dataset for each brain region using the end-to-end model. Shaded regions confidence intervals, with mean AUC values and standard deviations reported.

When compared to the baseline model results (Figure 13), the end-to-end model exhibited significant improvements, especially in PR curves, where RPFC activations demonstrated a clear advantage. The differences in precision and recall were more pronounced for RPFC, highlighting its robust performance under the end-to-end framework. While SMA showed comparable results across both models, RPFC activations—consistently the top performer in the end-to-end model—underperformed with the baseline model. This suggests that preprocessing in the baseline model may have obscured critical features in RPFC activations, reducing its effectiveness. Full performance metrics for the baseline model are included in the supplementary section.

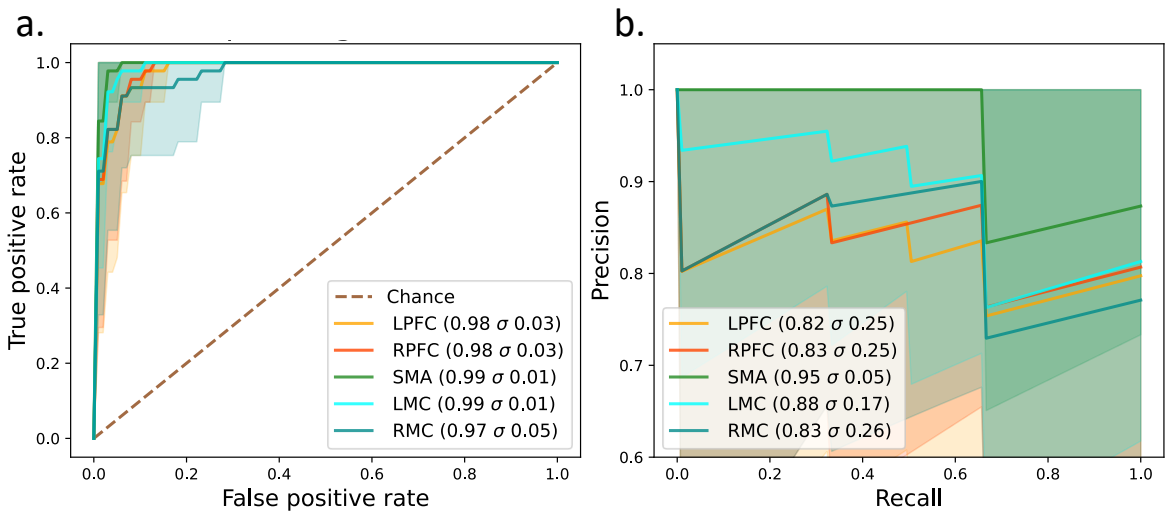

**Figure 13:** (a) ROC and (b) PR curves obtained from 15-fold cross-validation performed on the FLS-PC dataset for each brain region using the baseline model. Shaded regions denote confidence intervals, with mean AUC values and standard deviations reported.

## 3.4. Leave-one-subject-out (LOSO) cross-validation (CV)

We carried out LOSO CV to evaluate the misclassification error (MCE) for unseen subjects excluded from the training set. From Figure 14, the average MCE for all three tasks is less than 10%, with generally low variance, demonstrating the robustness of the end-to-end model.

For the FLS suturing task (Figure 14a), the MCE remains low across all regions, with minimal outliers. In contrast, for the ETI task (Figure 14b), two subjects - LC20, and LC21 - consistently emerge as outliers across all brain regions, displaying higher MCE (>25%). Although this suggested possible systematic differences specific to these individuals or issues related to data acquisition, issues weren't directly evident from the reinspection of the signals. A closer examination of task performance revealed that all trials of L20 were unsuccessful in the dataset, while all trials of L21 were incorrectly predicted as successful. This appears to be a case of bias, where the model perceives all brain activations of the held-out subject as similar, indicating no cortical changes associated with learning. However, this represents a 2 out of 20 error rate (10% misclassification), which could be further reduced by incorporating a broader cohort of individuals. For the FLS pattern cutting task (Figure 14c), the MCE is close to zero, and no outliers are observed, confirming strong model performance. The table with average classification metrics for LOSO CV is provided in the supplementary section.

Figures 14d, 14e, and 14f depict results from the baseline model for the same tasks. Compared to the end-to-end model, the baseline model exhibits higher MCEs and increased variance across all tasks. Notably, the regions that performed well in the end-to-end model (e.g., RPFC) show more pronounced errors in the baseline model, reinforcing the superior reliability and consistency of the end-to-end approach.

These results further highlight the capability of the end-to-end model to generalize effectively across unseen subjects and outperform the baseline model, even under challenging LOSO conditions. Full metrics for LOSO CV are provided in the supplementary section.

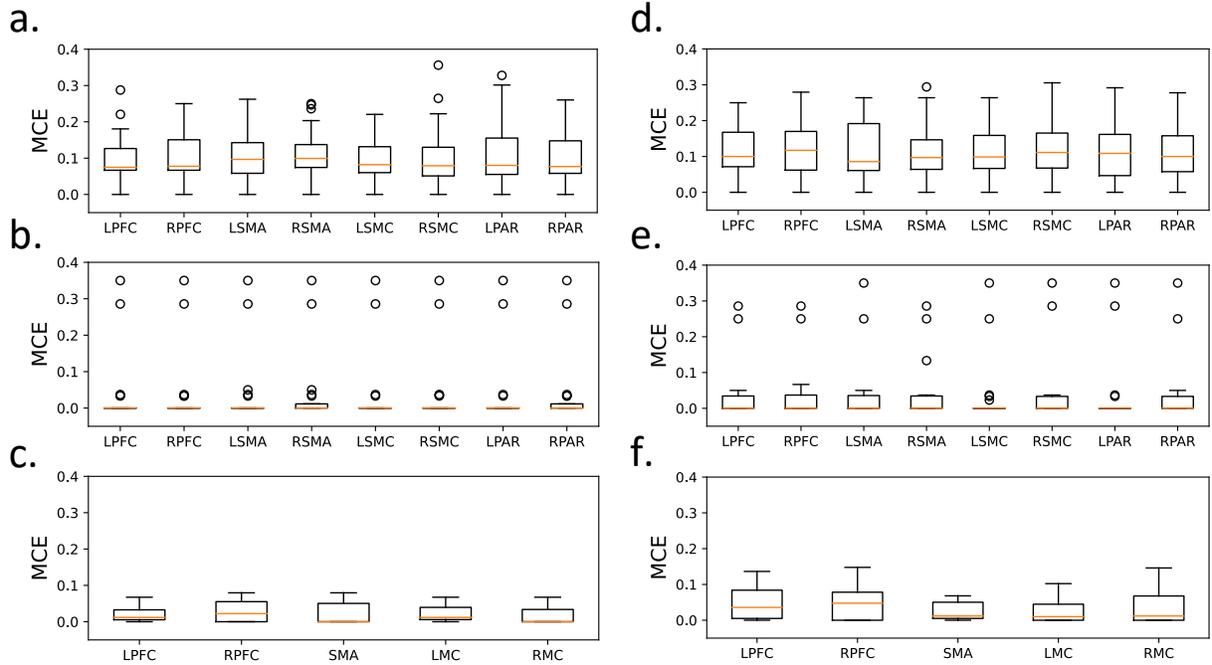

**Figure 14**: Boxplots showing the misclassification error (MCE) of test subject trials during leave-one-subject-out (LOSO) cross-validation using fNIRS signals from respective brain regions for the (**a**) FLS suturing (**b**) ETI (**c**) FLS pattern cutting tasks. Adjacent plots (**d**), (**e**) and (**f**) present the corresponding MCE results for the same tasks using the baseline model.

## 4. Discussions

The results demonstrate that neuronal activations from brain regions are capable of accurately distinguishing between expertise levels in surgery and emergency medicine. The accuracy of skill classification using the baseline model with fully processed brain imaging signals is statistically lower ($p < 0.05$) compared to that of end-to-end model with raw signals for all tasks except the ETI task. A one-sided Wilcoxon's signed rank test confirms that there is a significant difference between most of the performance metrics of the two classifiers, as evident from Table 9. The alternative hypothesis tested whether the baseline model delivers significant median improvement over end-to-end model. The one-sided Wilcoxon signed-rank test show that the baseline scores are not higher for any of the metrics, as seen in Table 9 in supplementary materials. This confirms that even for ETI task, using the baseline model does not provide a significant advantage.

**Table 9.** The p-values from the one-sided Wilcoxon signed-rank test for each task and performance metric. Statistically significant (p<0.05) results are in bold.

| Tasks | 15-fold Cross-validation Results | | | | | | | Test Set Results |
| --- | --- | --- | --- | --- | --- | --- | --- | --- |
| | MCC | F1-score | Sensitivity | Specificity | Accuracy | ROC-AUC | PR-AUC | Accuracy |
| **FLS_S** | **0.007** | **0.02** | 0.629 | **0.008** | **0.012** | **0.004** | **0.004** | |
| **FLS_SRet** | 0.23 | 0.273 | 0.23 | 0.191 | 0.23 | 0.055 | **0.012** | |
| **ETI** | 0.074 | 0.117 | 0.68 | **0.029** | **0.046** | 0.528 | 0.809 | 0.191 |
| **FLS_PC** | **0.031** | **0.031** | **0.03** | **0.031** | **0.031** | 0.051 | 0.062 | **0.031** |

The area under the PR and ROC curves in the Results section demonstrate that the classifiers separate successful and unsuccessful trials with high accuracy. The AUCs being independent of fixed cutoff values, provide a more comprehensive representation of model performance than the metrics like specificity and sensitivity. Another reliable metric independent of cutoff points is the trustworthiness metric [Wong] which is described in detail elsewhere [Yanik 2023]. The trustworthiness metrics for each test are given in the supplementary section (Table 6 through 8). The NetTrustScore greater than 0.8 for all the tests (Table 6 through 8 in supplementary materials) suggests that the end-to-end model classifies expertise level with high confidence and a greater margin.

In summary, based on the results for both training and testing schemes, brain activation from sensory motor regions consistently differentiates expertise levels with high accuracy across tasks, while also generalizing well to retention sets when using raw signals. Contralateral PFC activations exhibit different discriminative powers between the training set and the skill-retention set for complex suturing tasks; RPFC activations are more discriminative in the training set, while LPFC activations generalize better to skill retention. However, LPFC activations underperform compared to other regions in the pattern-cutting task, where RPFC and RMC activations are the most discriminative. A recent review paper by Andersen on regions of interest for surgical assessment highlights the observation of higher activations in left PFC regions among novices in many studies. The dorsolateral PFC is crucial for executive functions, with the left PFC being involved when individuals anticipate potential conflicts and the right PFC during the actual experience of conflict [Pulopulos]. Robinson has further shown, through lesion studies (n=90), that damage to the right PFC region results in deficits in autonomic suppression. Hence, the conflict resolution and suppression abilities to maintain focus facilitated by the right PFC appear to be important distinguishing markers of proficiency during learning.

For the ETI task, performance metrics across all brain regions were found to be similar. Additionally, it can also be observed that training set accuracy is mostly lower than retention set accuracy. This can be attributed to the nuanced differences in the task-specific brain functional connectivity that are still forming during the learning process but are consolidated when the trainee successfully retains the skills after sufficient practice.

Furthermore, we investigated the effectiveness of measured hemodynamic signals from the latter trials in a session, particularly considering the potential for fatigue and hemodynamic saturation. To assess this, we examined classifier performance when trained exclusively on first trial of each day, then the second trial and so on. To ensure fair comparison, we included only the trials from Day 10 onwards, as the 7-trial sessions begin from Day 11 for the FLS-S task. Balanced accuracy, measured as the average of specificity and sensitivity for each brain region, is observed in a stacked line graph in Figure 15.

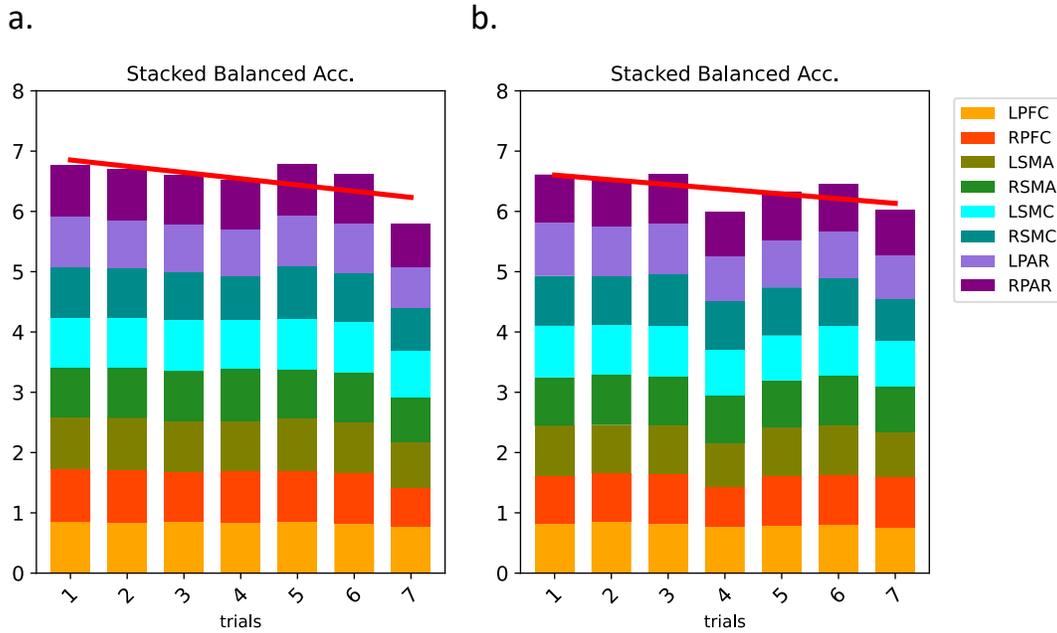

**Figure 15.** Stacked bar graphs obtained by stacking balanced accuracy metrics from all brain regions, for each trial separately. Performance during cross validation on training set (a) and on test set (b). Trendline is a first order polynomial fit to the sums.

Additionally, the study reveals the influence of session duration on model performance. Classifier performance shows a slight decline after the third trial in extended sessions, likely due to neurovascular coupling (NVC) attenuation from prolonged activation. NVC mechanisms, which regulate cerebral blood flow (CBF) through components like astrocytes, vascular smooth muscle cells (VSMCs), and pericytes, could lead to saturation in blood HbO concentration changes with extended activation periods [Devor, Kisler, Attwell]. The observed dip in test set performance, particularly in later trials, suggests that this attenuation might impact the model's ability to generalize when trained on data from extended sessions. This insight underscores the importance of considering the temporal aspects of data collection when developing models for real-time skill assessment using fNIRS signals. Our analysis indicates that these hemodynamic shifts in the data stream are not adequately addressed in current processing methods, which, as shown in this study, clearly affect performance. By focusing on the raw signal data, the end-to-end model demonstrated robustness against such issues, further highlighting the advantages of processing raw fNIRS signals directly rather than relying on heavily preprocessed data that might obscure important neural information.

## 5. Conclusion
The end-to-end deep learning model developed in this study demonstrates a high degree of accuracy in distinguishing the proficiency levels of surgical trainees and emergency medical practitioners. By leveraging raw neuronal activations measured using a portable fNIRS device, the model eliminates the need for extensive preprocessing and feature extraction, offering a streamlined approach to skill assessment. The model achieved a mean classification accuracy of 93.9% (SD 4.5) and a mean generalization accuracy of 90.5% (SD 3.5) on an unseen skill retention dataset. Furthermore, using leave-one-subject-out cross-validation, the model classified proficiency levels with a mean accuracy of 94.1% (SD 3.6), underscoring its robustness and adaptability.

This work establishes a framework for real-time, non-invasive assessment of bimanual motor skills in medical and surgical training environments. However, certain limitations underscore opportunities for future research and development. A key limitation is the lack of interpretability techniques specifically designed for time-series data, which hinders a comprehensive understanding of the model's decision-making processes. While methods such as Grad-CAM have been effectively used in spatial domains [Chollet] and have been adapted to time-series data in some cases [Fawaz], their application to time-series data with varying input sizes warrants further investigation.

Future research could focus on integrating methods that enhance interpretability frameworks into the deep learning pipeline, allowing for more transparent insights into cortical activations during motor skill tasks. Additionally, there is significant potential in adopting large language model (LLM)-like architectures to analyze extensive datasets of cortical activations, potentially improving generalization across different tasks and scenarios. Expanding the scope of this research to include a broader range of tasks and exploring how specific functional patterns relate to cortical activations could provide richer insights. A more granular investigation of neural activity beyond the eight brain regions studied here could also yield novel insights into the neural mechanisms of skill learning and retention.

While the model demonstrates efficacy in controlled environments, future research should aim to integrate metrics such as stress and fatigue, which are pivotal in emulating real-world clinical scenarios. These factors significantly influence neuronal activations and task performance, making their inclusion essential for ensuring the model's robustness under varying physiological and environmental conditions. Incorporating these variables would enhance the model's applicability in high-stakes, real-world scenarios, enabling more accurate and comprehensive performance evaluations. Such enhancements are particularly important for clinical training, where stress and fatigue are inherent and unavoidable, and their effects on skill execution must be accounted for.

Beyond its immediate applications in medical and surgical training, this framework offers substantial potential for broader use in domains requiring precise motor skill assessment, such as robotics, sports training, and rehabilitation. Its adaptability across diverse contexts underscores the versatility of leveraging neuronal activations as biomarkers for performance evaluation. Additionally, ensuring real-time applicability necessitates investigations into computational efficiency and latency to assess the feasibility of deployment in live training systems. Integrating adaptive feedback mechanisms could further enhance its utility by providing trainees with immediate, actionable insights, facilitating targeted skill improvement. Addressing challenges such as variability in data quality due to differences in device placement or signal noise will also be critical for practical implementation. By resolving these aspects, the model can evolve into a comprehensive tool for skill assessment and training, paving the way for personalized, effective methodologies across multiple fields.

**Acknowledgements**

The authors gratefully acknowledge the support of this work through the Medical Technology Enterprise Consortium (MTEC) award #W81XWH2090019, and the U.S. Army Futures Command, Combat Capabilities Development Command Soldier Center STTC cooperative research agreement # W912CG2120001.